# Controlled Fabrication of Uniform Columnar Porous GaN Architectures by Electrochemical Etching

Thom R. Harris-Lee,[a*] Ben Thornley,[a] Jiawei Zhang,[a] Martin Frentrup,[a] Simon Fairclough,[a] Menno J. Kappers[a] and Rachel A. Oliver[a]

[a] Department of Materials Science and Metallurgy, University of Cambridge, Cambridge, CB3 0FS, UK

*trh51@cam.ac.uk

## Abstract

Electrochemical etching (ECE) of *n*-type *c*-plane GaN enables the formation of porous architectures capable of enhancing the performance of optoelectronic devices and broadening the range of possible GaN applications. Here, we demonstrate that the addition of NaOH to $Na_2C_2O_4$ (and other non-acidic etchant solutions) directs ECE toward the formation of nanoscale, vertically aligned, highly uniform columnar pore morphologies with relatively limited surface roughening. By varying the NaOH concentration within mixed electrolyte systems, the columnar pore shape, diameter, spacing, and bulk and surface porosity can be readily tuned. In addition to manually mixing NaOH with another etchant, aqueous basic electrolytes such as $Na_2CO_3$ and $Na_3PO_4$, each inherently containing significant hydroxide concentration through hydrolysis, also produce similar columnar pore morphologies, indicating that hydroxide plays the key role in directing the vertical pore growth. Chronoamperometry and pore structure analysis reveals that the mixed hydroxide-containing electrolyte systems exhibit a distinct, new ECE mechanism rather than a simple superposition of the individual ECE behaviour of each anion present. This work provides significant insight into the ECE mechanism for porous GaN, which is proposed to contain multiple reaction steps where hydroxide anions are involved in the formation of a reactive intermediate. These findings establish a versatile and controllable route for engineering columnar porous GaN architectures, offering significant advancement in applications requiring high porosity, ordered, and bespoke structures.

## Introduction

Gallium nitride (GaN) is a wide-bandgap (3.4 eV) III-V semiconductor known for its high thermal stability, chemical resistance, and excellent electronic and optoelectronic properties, being widely used in efficient LED lighting and high-power transistors.[1–3] Porous GaN has emerged as a promising component in overcoming common issues associated with GaN-based devices such as difficult strain management.[4,5] The availability of porous GaN architectures also broadens the range of possible applications of GaN, facilitating catalysis[6,7] and sensing[8,9] (due to the high surface area), and composite formation by pore infiltration.[10,11] Electrochemical etching (ECE) is a conductivity-selective process that can be applied to *n*-type GaN to produce porous GaN layers by an anodic electrochemical reaction (Equation 1).

$$2GaN + 6h^{+} \rightarrow 2Ga^{3+} + N_2 \qquad (1)$$

Holes ($h^+$) may be generated in the *n*-type GaN by an applied bias potential in two possible ways, by Zener tunnelling or by impact ionisation.[12] These holes then oxidise GaN to result in aqueous $Ga^{3+}$ and gaseous $N_2$ products, although the mechanistic steps between hole-generation and product formation in Equation 1 are not known and often disputed.[13–16]

Pore morphology within a porous architecture can be readily tuned by a range of ECE conditions, most prominently: applied bias potential, electrolyte composition, and *n*-type GaN free carrier density.[12,14] Previous studies have demonstrated the significant influence that different electrolyte compositions have on the resulting pore morphology, even under otherwise similar ECE conditions.[14,15,17–19] This suggests that the etchant anion plays a key role in the ECE mechanism, with different anions resulting in different preferential removal of GaN, giving rise to distinct pore morphologies. Despite this variation, the use of mixed electrolytes, for example to attempt to combine the features of two different pore morphologies, has received little attention.[14,15,20] We have previously shown the use of pH buffer solutions to stabilise pH during ECE, and within that same work explained that many active anions may be present from a single added electrolyte chemical, effectively resulting in a mixed electrolyte system.[14] The potential of mixed electrolytes to enable new and more tuneable pore morphologies remains largely unexplored.

A key limitation of many existing porous GaN structures is the trade-off between porosity and useful resulting porous architecture. High porosity is often accompanied by non-uniform ECE and poor surface integrity, while more ordered pore structures with smoother surfaces suitable for overgrowth or device integration often exhibit relatively low porosity.[14,21] Here, we introduce a new and controllable ECE strategy based on the addition of hydroxide ($OH^-$) anions to a non-acidic electrolyte system to form uniform, highly porous, vertically-aligned columnar pore architectures, whilst retaining the integrity of the surface.

## Results and Discussion

### Columnar Pore Formation

A wide range of electrolytes have been reported to successfully electrochemically etch *n*-type GaN, leading also to a number of different possible pore morphologies. While the specific changes in the ECE mechanism caused by the different etchant species are currently unclear, it has been proposed that these resulting pore morphology differences may arise from variations in the binding affinity of etchant anions to $Ga^{3+}$ cations generated during the ECE process.[14,19] To further investigate these possibilities, NaOH and $Na_2C_2O_4$ were combined into a mixed etchant solution (0.1 M total electrolyte concentration in a 50 ml solution, containing 0.05 M NaOH and 0.05 M $Na_2C_2O_4$). By mixing these two etchants, it is possible that a new porous architecture will result, combining the best and complementary characteristics of each individual etchants ECE behaviour. This approach also enables assessment of whether the resulting ECE process reflects both etchant mechanisms occurring simultaneously, one etchant mechanism dominating over the other, or if an entirely new, distinct ECE mechanism results.

First, considering each etchant individually, ECE performed in NaOH at a bias potential of 4 $V_{Ag/AgCl}$ results in an overall vertically-aligned pore morphology, although the uppermost ~130 nm exhibits a slightly different structure containing a more branched, lateral pore morphology (Figure 1a). The surface of the etched sample appears faceted and non-uniform, displaying regions of hexagonal surface pits and large areas of little or no apparent ECE occurrence (Figure 1d). In contrast, ECE using $Na_2C_2O_4$ at 8 $V_{Ag/AgCl}$ produces a highly porous structure but with poor uniformity and limited control, accompanied by an extremely rough surface morphology (Figure 1b, Figure 1e). Different applied potentials for each etchant were chosen for this comparison due to the different ECE behaviour of each. ECE with 0.1 M NaOH at applied potentials higher than 4 $V_{Ag/AgCl}$ resulted in a very unusual cross-sectional morphology (Figure S2a), appearing as if the sample no longer has structural integrity, and ECE with 0.1 M $Na_2C_2O_4$ at lower applied potentials than 8 $V_{Ag/AgCl}$ took impractically long durations (Figure S2b). It is noted that ECE in NaOH solutions initially appeared to result in inconsistent pore morphologies when experiments were repeated. However, when NaOH solutions were prepared immediately before the ECE experiment, the results became more consistent (see SI, Figure S2a,c,d for a comparison of ECE in NaOH solutions prepared at different times, all at 8 $V_{Ag/AgCl}$). This difference is therefore likely due to the impurity of the as-purchased NaOH pellets (<2% carbonate), reaction of the NaOH pellets with $CO_2$ from air before dissolution, and/or the dissolution of $CO_2$ from air into pre-made solutions over time, all resulting in the presence of carbonate anions ($CO_3^{2-}$) within the etchant solution, which, as will be discussed later, will significantly influence the ECE process. It is therefore strongly advised that all solutions used for ECE should be handled carefully and freshly prepared for each experiment to limit irreproducibility.

ECE conducted at 8 $V_{Ag/AgCl}$ in the mixed etchant solution results in a highly vertical, columnar pore morphology that appears to combine the strong vertical alignment characteristic of NaOH ECE with the high porosity typically achieved when using $Na_2C_2O_4$ (Figure 1c). The resulting film surface is also more porous than for NaOH alone, but significantly less roughened than $Na_2C_2O_4$ alone (Figure 1f). However, chronoamperometry data (etching current vs time, where current is normalized to the original planar geometric surface area of the etched sample region) shown in Figure 1g,h indicates that ECE in NaOH at 4 $V_{Ag/AgCl}$ proceeds almost 100 times faster than ECE in $Na_2C_2O_4$ at 8 $V_{Ag/AgCl}$ (~40 s vs ~3750 s, respectively). Given that the mixed solution ECE was performed at 8 $V_{Ag/AgCl}$, which for NaOH alone was deemed too high a voltage, and given that ECE completed in just ~100 s (SI, Figure S2a), it would be expected that the NaOH component of the mixed solution would dominate the ECE behaviour, yielding a morphology similar to that produced by NaOH at 8 $V_{Ag/AgCl}$. Since this does not occur, it cannot be the case that only NaOH-driven ECE occurs, nor could NaOH ECE occur followed by a more restricted $Na_2C_2O_4$-driven ECE. Instead, the formation of a new columnar pore structure, together with an intermediate ECE duration of ∼1000 s (Figure 1g,h), provides strong evidence that the mixture of both $OH^-$ and $C_2O_4^{2-}$ anions induces a new ECE mechanism, rather than the two independent mechanisms occurring simultaneously. The ECE process must therefore be a two-step (or more) mechanism, where the $OH^-$ anions are more effective in one step, and the $C_2O_4^{2-}$ anions are more effective in the other step, resulting in a new mechanism that combines the favoured properties of each etchant species.

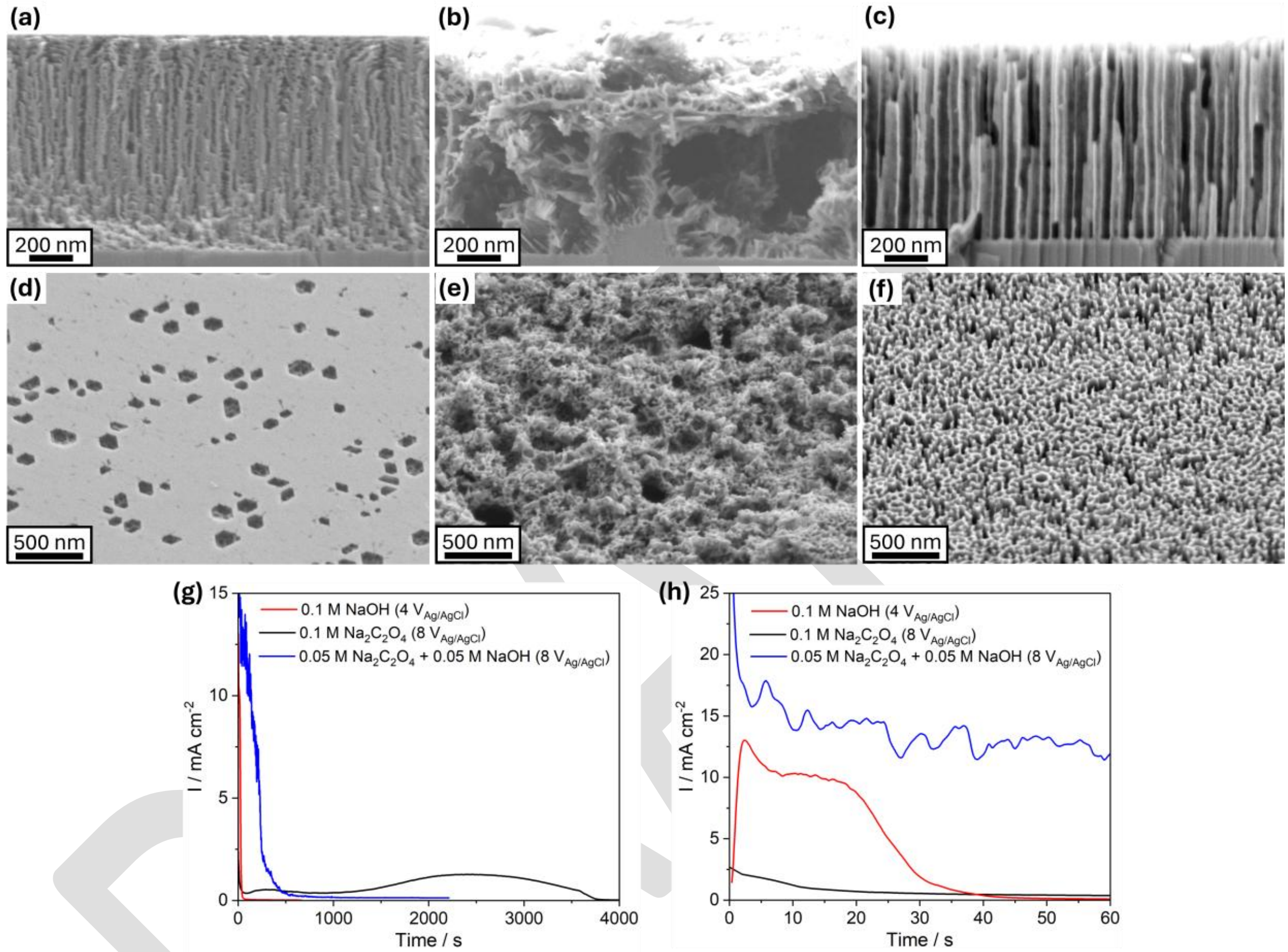

**Figure 1.** SE cross-section images and 45° tilted SE plan-view images for 1 µm *n*-type GaN ([Si] = 1.5 × $10^{19}$ $cm^{-3}$) electrochemically etched in (a,d) 0.1 M NaOH at 4 $V_{Ag/AgCl}$, (b,e) 0.1 M $Na_2C_2O_4$ at 8 $V_{Ag/AgCl}$, (c,f) mixed 0.05 M NaOH and 0.05 M $Na_2C_2O_4$ at 8 $V_{Ag/AgCl}$. (g) Chronoamperometry data from each ECE process, (h) chronoamperometry of only the first 60 s of ECE.

### ECE Mechanistic Interpretation

No direct evidence of the ECE mechanism is obtained in this study. Nevertheless, the results obtained using the mixed NaOH-$Na_2C_2O_4$ system, together with comparison with previously reported electrolyte systems, provide useful insight into the factors that may govern the ECE process. The mechanistic interpretation proposed here is therefore necessarily speculative and is based on consistent trends observed across a broad range of ECE conditions, rather than on direct measurement of the proposed mechanistic processes.

Oxalate $C_2O_4^{2-}$ anions would be expected to bind more strongly to $Ga^{3+}$ than $OH^-$ anions due to their chelating nature, therefore the final step of a mechanism that involves removal of $Ga^{3+}$ cations into a solution of $OH^-$ and $C_2O_4^{2-}$ anions is very likely to result in $Ga^{3+}$ coordinated only to $C_2O_4^{2-}$. In this case, assuming the simplest scenario of a two-step mechanism (after electrochemical oxidation has occurred), the contribution of $OH^-$ anions to the mechanism in the mixed solution must mean that $OH^-$ will mediate the first step to form a reactive intermediate. $C_2O_4^{2-}$ will then dominate the second step to fully remove the $Ga^{3+}$ ion from the GaN lattice. The fact that $OH^-$ is more favoured towards forming the reactive intermediate may be key to understanding the formation of columnar pore morphologies. For example, if the formation of the $OH^-$-mediated reactive intermediate accelerates the ECE reaction, as observed by chronoamperometry, then the geometric effect of the electric field being concentrated at pore tips may have a greater influence on the resulting pore morphology (i.e., ECE becomes less reaction-limited and more hole-generation/GaN oxidation-limited).[16]

While this mechanistic interpretation is currently speculative and is the subject of ongoing investigation, it is consistent with empirical trends we have reported previously describing the influence of electrolyte composition on pore morphology.[14] It should be noted that variations in applied potential and doping density has a significant impact on the resulting pore morphologies for a given electrolyte[21,22]. Assessing the universality of the electrolyte groupings proposed below would therefore require similarly systematic and comprehensive studies across a range of doping densities and applied potentials, to determine whether the different morphologies that may emerge under these conditions can nevertheless be classified within the same electrolyte groupings identified here and in previous work[14]. From the conditions chosen here and in our earlier work, resulting pore morphologies can be grouped to electrolyte properties as follows (summarised in Table 1, all ECE was specifically performed at 8 $V_{Ag/AgCl}$ with 1 µm *n*-type GaN, [Si] = $1.5 \times 10^{19}$ $cm^{-3}$): (i) acidic and neutral electrolytes containing weakly coordinating ligands all produce similar, low porosity, laterally-oriented pore morphology (NaCl, HCl, $HNO_3$, $NaNO_3$, $CH_3COOH$, $CH_3COONa$) branching from an electrochemically etched threading dislocation; (ii) acids capable of forming strong, chelating ligands by multi-step dissociation produce branching, 'wavey' pore morphologies ($H_2C_2O_4$, $H_2SO_4$, $H_3PO_4$); (iii) neutral or only mildly basic (i.e., negligible concentration of $OH^-$ anions) conjugate salts of the weak acids in group (ii) result in aggressive, destructive ECE behaviour ($Na_2C_2O_4$, $Na_2SO_4$); (iv) more strongly basic (i.e., significant concentration of $OH^-$ anions) electrolytes, forming strong, chelating anions, produce columnar pore morphologies ($Na_3PO_4$, as well as $Na_2CO_3$ and NaOH-$Na_2C_2O_4$ from this work).

**Table 1.** Pore morphologies grouped by electrolyte properties using empirical results from this work and previous work.[14] Note that all ECE was performed on 1 µm *n*-type GaN ([Si] = $1.5 \times 10^{19}$ $cm^{-3}$) with a bias potential of 8 $V_{Ag/AgCl}$. Different morphologies are possible with different applied bias potentials and dopant densities, hence this is not claimed to be a universal grouping system for any ECE condition except those used here.

| Group | Electrolyte Behaviour | Anions | Pore Morphology | Step 1 Mediator | Step 2 Mediator |
|---|---|---|---|---|---|
| (i) | 'Spectator' anions | $NO_3^-$, $Cl^-$, $CH_3COO^-$ | Low porosity | $H_2O$ | $H_2O$ |
| (ii) | Multi-step dissociation | $HC_2O_4^-/C_2O_4^{2-}$, $HSO_4^-/SO_4^{2-}$ | Branched | $H_2O$/electrolyte anion | Electrolyte anion |
| (iii) | Strong ligands | $C_2O_4^{2-}$, $SO_4^{2-}$ | Destructive | $H_2O$/electrolyte anion | Electrolyte anion |
| (iv) | Mixed $OH^-$ and strong ligand | $OH^-/CO_3^{2-}$, $OH^-/PO_4^{3-}$, $OH^-/C_2O_4^{2-}$ | Columnar | $OH^-$ | Electrolyte anion |

These groupings can be rationalised by the proposed two-step mechanism. For group (i), electrolyte anions are effectively inactive, as coordination of $H_2O$ to $Ga^{3+}$ is more favoured,[23–25] hence both the formation of the reactive intermediate (step 1) and subsequent $Ga^{3+}$ removal (step 2) is mediated by $H_2O$. For group (ii), strongly chelating ligands are accessible, but their formation is limited by their preference for the partially dissociated state.[14] Notably, these systems all uniquely exhibit current oscillations in chronoamperometry measurements, possibly due to this dissociation equilibrium shifting. Under these conditions, the initial step may be mediated by $H_2O$ or any of the possible electrolyte anions from different steps of dissociation, whereas the second step is likely predominantly mediated by the fully dissociated anion. For group (iii), complete dissociation leads to the presence of only strongly coordinating, chelating anions. Here, the first step may be mediated either by $H_2O$ or the electrolyte anion, while the second step mediated by the electrolyte anion. Finally, for group (iv), strong, chelating anions are again present, but now in combination with significant $OH^-$ concentrations. From the earlier discussion, the first step is likely now mediated by $OH^-$, while the second step remains mediated by the chelating anion. It should be emphasised that this mechanism has not been directly observed experimentally. Rather, it is a hypothetical mechanism derived from consistent correlations between electrolyte composition, chronoamperometry traces, and resulting pore morphologies.

**Columnar Pore Morphology Control: Influence of the $NaOH/Na_2C_2O_4$ Ratio**

The ratio of NaOH to $Na_2C_2O_4$ was systematically varied to assess the degree of control over columnar pore diameter, morphology, and spacing (i.e., pore wall width). Figure 2 shows SE cross-section, plan-view and AFM images for a range of NaOH molar percentages within a total 0.1 M mixed NaOH-$Na_2C_2O_4$ etchant solution. Corresponding chronoamperometry data and 45° tilted SE plan-view images are provided in the SI, Figure S3. $Na_2C_2O_4$ is very weakly basic, so hydrolysis of $C_2O_4^{2-}$ to form $OH^-$ is negligible ($pK_{b2}$=10.19 [26], Equation 2).

$$C_2O_4^{2-} + H_2O \rightleftharpoons HC_2O_4^- + OH^- \quad (2)$$

For a 0.1 M aqueous $Na_2C_2O_4$ solution, the hydrolysis reaction will form ~2.5 μM of $OH^-$ anions, as well as ~2.5 μM of $HC_2O_4^-$ anions. Since there is a negligible concentration of $OH^-$, and there is the same negligible concentration of $HC_2O_4^-$ anions (which would be required for the reverse reaction in Equation 2 to occur upon addition of NaOH), the concentration of NaOH added to the mixed solution is effectively equal to the total concentration of $OH^-$ anions present. It is therefore easy to control the ratio of $OH^-/C_2O_4^{2-}$ anions within this mixed electrolyte for a systematic study on the influence of the mixture composition on the ECE mechanism.

Herein, the mixed NaOH-$Na_2C_2O_4$ solutions will be denoted by the molar percentage of NaOH, given that the total concentration will always be 0.1 M; for example, a mixed NaOH-$Na_2C_2O_4$ solution consisting of 0.025 M NaOH and 0.075 M $Na_2C_2O_4$ will be referred to as the '25%-NaOH' solution. Importantly, despite the earlier discussion on the observed irreproducibility of NaOH electrolyte alone, the columnar morphologies produced across the $NaOH/Na_2C_2O_4$ composition range were highly reproducible and showed no apparent dependence on the age of the NaOH pellets or previously prepared NaOH solution used in the electrolyte preparation. The solution resistances between two platinum electrodes with 2 cm separation and pH values of the bulk electrolyte mixtures were measured before ECE (solution resistances of 57.2, 59.1, 57.1, 54.7, 56.0, 61.7, 53.3, 55.4, and 48.4 Ω and pH values of 8.4, 12.0, 12.4, 12.6, 12.7, 12.8, 12.9, 13.0, and 13.0 were measured for the solutions 0.1 M $Na_2C_2O_4$, 10%-NaOH, 25%-NaOH, 40%-NaOH, 50%%-NaOH, 60%-NaOH, 75%-NaOH, 90%-NaOH, and NaOH, respectively). The solution resistances were all on a comparable scale, particularly when compared to the relatively high resistance from the GaN electrode (on the scale of kΩ, measured by electrochemical impedance spectroscopy), and the measured pH values of each bulk solution matched the expected values based on the manually added concentration of NaOH. Following ECE, no measurable change in the bulk solution resistance or pH (at the resolution of the pH meter, 0.1 pH units) was observed. This does not imply that the electrolyte composition at the reaction surface remained unchanged during ECE; rather, any such local changes cannot be resolved using conventional bulk measurements.

Figure 2 presents cross-sectional SE images alongside corresponding images of each surface, recorded by AFM in all cases except for the sample etched in 10%-NaOH (Figure 2a), where the high surface roughness necessitates the use of SE imaging. Figure 2a shows that ECE at 8 $V_{Ag/AgCl}$ in the 10%-NaOH solution produces a vertically-oriented pore morphology, but with substantial volumes of the *n*-type GaN layer removed and a severely roughened surface. However, for 25%-NaOH and greater solutions, (up to 90%-NaOH), well-defined, uniform columnar pore morphologies are consistently formed across the entire layer. The surface pores from ECE in 25%-NaOH and 40%-NaOH appear interconnected (as did 50%-NaOH from Figure 1f), while the surface pores from 60%-NaOH, 75%-NaOH, and 90%-NaOH are all spatially isolated. The RMS roughness, measured over the 1 $\mu m^2$ AFM scan areas shown in Figure 2, varies systematically with the relative NaOH/$Na_2C_2O_4$ ratio, decreasing from 11.4 nm at 25% NaOH to 3.4 nm at 60% NaOH, where it reaches a minimum. The roughness remains unchanged at 75% NaOH (3.4 nm) before increasing gradually to 4.8 nm at 90% NaOH (RMS roughness values of 11.4, 8.8, 3.4, 3.4, and 4.8 nm for 25%-NaOH, 40%-NaOH, 60%-NaOH, 75%-NaOH and 90%-NaOH, respectively).

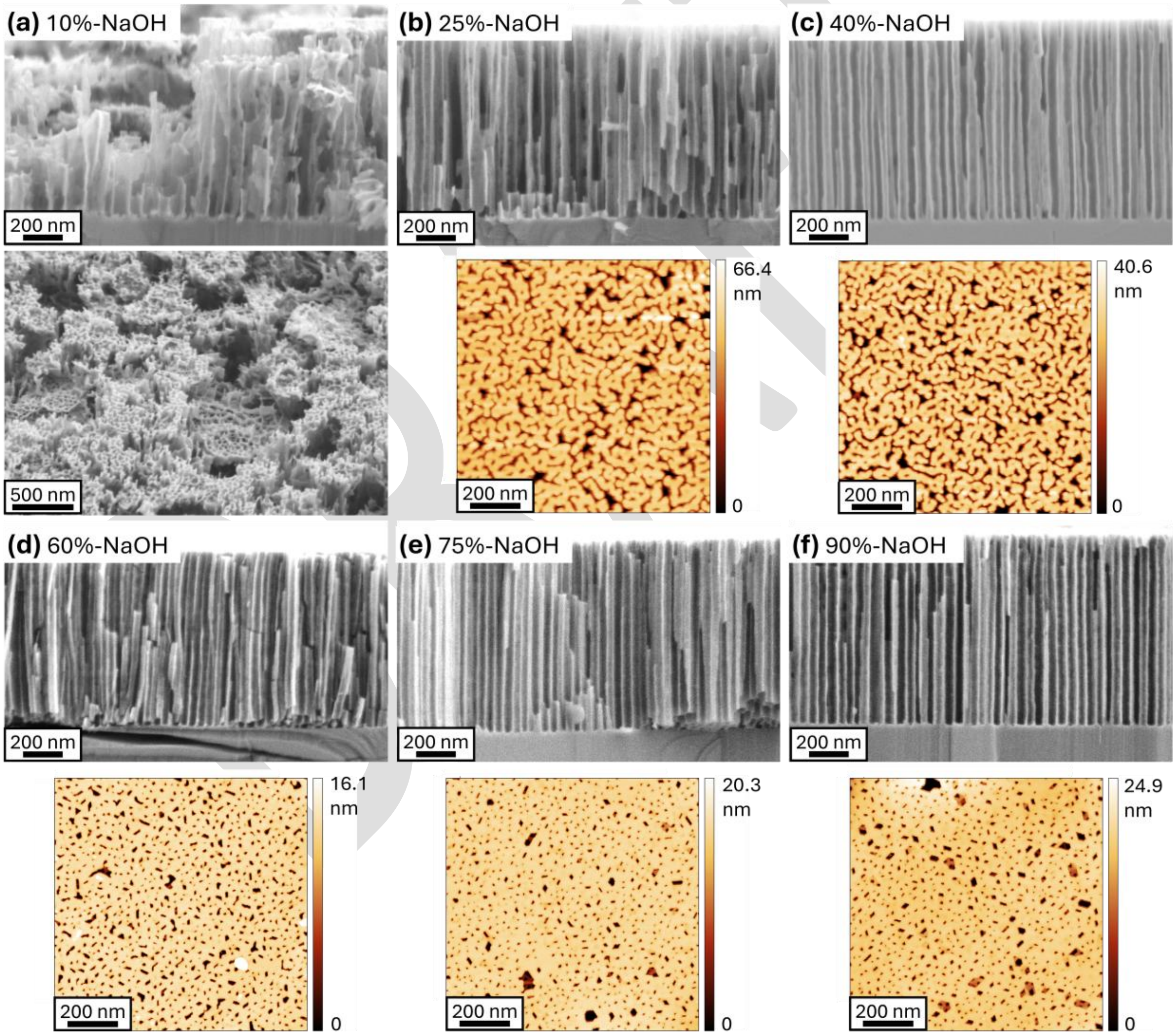

**Figure 2.** SE cross-section (top) and SE 45° tilted or AFM (bottom) surface images for 1 µm *n*-type GaN ([Si] = 1.5 × $10^{19}$ $cm^{-3}$) electrochemically etched at 8 $V_{Ag/AgCl}$ in a mixture of (a) 0.09 M $Na_2C_2O_4$ and 0.01 M NaOH, (b) 0.075 M $Na_2C_2O_4$ and 0.025 M NaOH, (c) 0.06 M $Na_2C_2O_4$ and 0.04 M NaOH, (d) 0.04 M $Na_2C_2O_4$ and 0.06 M NaOH, (e) 0.025 M $Na_2C_2O_4$ and 0.075 M NaOH, (f) 0.01 M $Na_2C_2O_4$ and 0.09 M NaOH.

Each columnar morphology from Figure 2b-f appears qualitatively similar, hence it is important to analyse the morphology of each in more detail to be able to evaluate the effect of NaOH composition in the etchant solution, as well as identify the optimal structure to apply to a specific application. Each individual columnar pore in Figure 2b-f appears uniform and relatively unchanging through the entire porous layer, hence the overall pore and pore wall properties could be extracted from a single plan-view image of the structure from within the 1 μm-thick porous layer. The data obtained by analysis of this is then assumed to be representative of the entire subsurface porous layer. This cannot be achieved by simple plan-view SE imaging or AFM as the sample surface appears distinct from the rest of the bulk region, an observation that is ubiquitous across all porous GaN morphologies.[14,15,27] Figure 3a-f contains plan-view SE images for columnar pore morphologies, where the uppermost 500 nm of the porous GaN layer has been removed by FIB milling (described in detail in the Methods section), revealing the bulk structure. Qualitatively distinct columnar morphologies are now evident for the samples etched in varying molar percentage of NaOH in the mixed NaOH-$Na_2C_2O_4$ solution. Quantitative porous structural data (porosities, pore areas, pore wall thicknesses, pore densities) has been obtained by analysis of these plan-view images (Figure 3g-i); detailed explanation of the image processing to obtain these values is included in the Methods section. The sample electrochemically etched in the 90%-NaOH electrolyte appears to show significant pore merging. However, no cross-sectional images of this sample gave any clear evidence of this occurrence, hence it is possible that this is a result of damage during the FIB milling procedure. Analysis of this plan-view image is therefore performed in a selected region with minimal damage (within the annotated red box, Figure 3f).

Figure 3g shows that the porosity of the samples as a function of NaOH proportion in the electrolyte varies greatly, with the lowest value of 15.5% for 25%-NaOH, and highest value of 59.9% for 50%-NaOH (Figure 3g). There is an initial trend of increasing porosity with increasing percentage of NaOH, up to a peak at 50%-NaOH, after which the porosity initially drops considerably to 41.1% for 60%-NaOH, remains similar for 75%-NaOH, and finally increases significantly again for 90%-NaOH. The pore density (number of pores per $\mu m^2$) consistently appears roughly inverse to the porosity. As may be expected, the average pore area trend matches that of porosity. Generally, smaller pores show higher pore density (Figure 3g), suggesting that the pore formation is driven by the amount of available space, and the pore area (lateral reach) may be controlled by the comparative rates of pore initiation and lateral growth. Pore wall thickness measurements (Figure 3i, in which the mean values are shown as red squares) are also consistent with expected trends based on pore density and pore areas; increasing mean pore area results in decreased pore wall thickness, and the wall thickness of 60%-NaOH and 75%-NaOH is lower than that for 40%-NaOH because, while all three possess similar mean average pore area, the pore density of 60%-NaOH and 75%-NaOH is greater.

Figure 3h and Figure 3i also show the distributions of pore sizes and pore wall thicknesses resulting from each etchant composition, displayed as box plots. For samples in both Figure 3h and Figure 3i, except for 90%-NaOH, there is no significant deviation between the mean and median values, and both these values are reasonably centred within the interquartile range, hence the distributions of pore area/pore wall thickness are relatively symmetrical. In contrast, the 90%-NaOH electrolyte possesses significant separation between the mean and median values for pore area, indicating that the distribution is skewed by unusually large pore area values. This is likely due to the merged pores (which, as noted above, may be due to FIB damage) being too difficult for the watershed algorithm to separate (Figure S1a). We note that this is the case despite the calculations performed here being undertaken on an area with relatively little pore merging, to limit the impact of any possible FIB damage on our analysis.

The 50%-NaOH solution results in the largest distribution of pore areas, while 25%-NaOH results in the least. In support, the 50%-NaOH sample shows a substantially larger standard deviation for pore area than the other etchant compositions (754 $nm^2$ for 50%-NaOH compared to 88, 296, 280, 237 $nm^2$ for 25%-, 40%-, 60%-, and 75%-NaOH, respectively). From the SE image in Figure 3c, it first appears that this may be due to the merging/overlap of multiple pores; however, given that there is no expected interconnection between pores (based on SE cross-sections measured within this study and from FIB-SEM tomographs reported previously for porous GaN-based distributed Bragg reflectors[28]), these merged pores are suggested to be a result of the FIB milling, and were therefore separated by watershed segmentation for all analysis performed, as shown in Figure S1. The large standard deviation remaining despite this correction indicates that the broad pore size distribution is an effect specifically of the 1:1 $NaOH/Na_2C_2O_4$ ratio. Assuming that surface pore nucleation sites are located randomly, and that the immediate subsurface lateral pore growth is uniform, it is possible that pore size distributions could be correlated to how dense the surface pores of a sample are – determining the possible variation in lateral growth before meeting another expanding pore and stopping. However, this would mean that the pore size distributions shown in Figure 3h should be greatest when using 90%-NaOH, and decrease with decreasing NaOH proportion, as this follows most sparse to least sparse surface pore distribution (see the AFM images in Figure 2). However, the pore area distribution for 40%- and 50%-NaOH are higher than both 60%- and 75%-NaOH, suggesting that this may not be the case, and that this observation is instead intrinsically linked to the ratio of etchant ions.

Interestingly, the variability in pore area is not mirrored in the pore wall thickness distributions (Figure 3i) or standard deviations, which remain reasonably similar for 40%-, 50%-, 60%-, 75%-, and 90%-NaOH mixtures (5.3, 4.1, 3.9, 3.9, 3.8 nm, respectively). This suggests that in these cases, pore wall thickness may be governed by a space-filling mechanism, where a pore grows laterally until its space-charge region overlaps with that of a neighbouring pore. Under a given set of ECE conditions, and given that all the pore shapes within a single columnar porous GaN sample appear very similar (Figure 2, Figure 3), the space-charge region width is expected to be relatively uniform for every pore, leading to a characteristic and consistent separation between pores, regardless of their individual size.[29] In contrast, 25%-NaOH displays the smallest standard deviation in pore area (~88 $nm^2$), but the largest in wall thickness (~8.3 nm). This behaviour suggests that the lateral pore growth in this regime may not be strictly limited by neighbouring space-charge regions overlapping, as it was for the other samples. Instead, pore area may be constrained intrinsically by the ECE mechanisms resulting from this ratio of anions, giving rise to less regulated spacing between pores but very consistent pore areas.

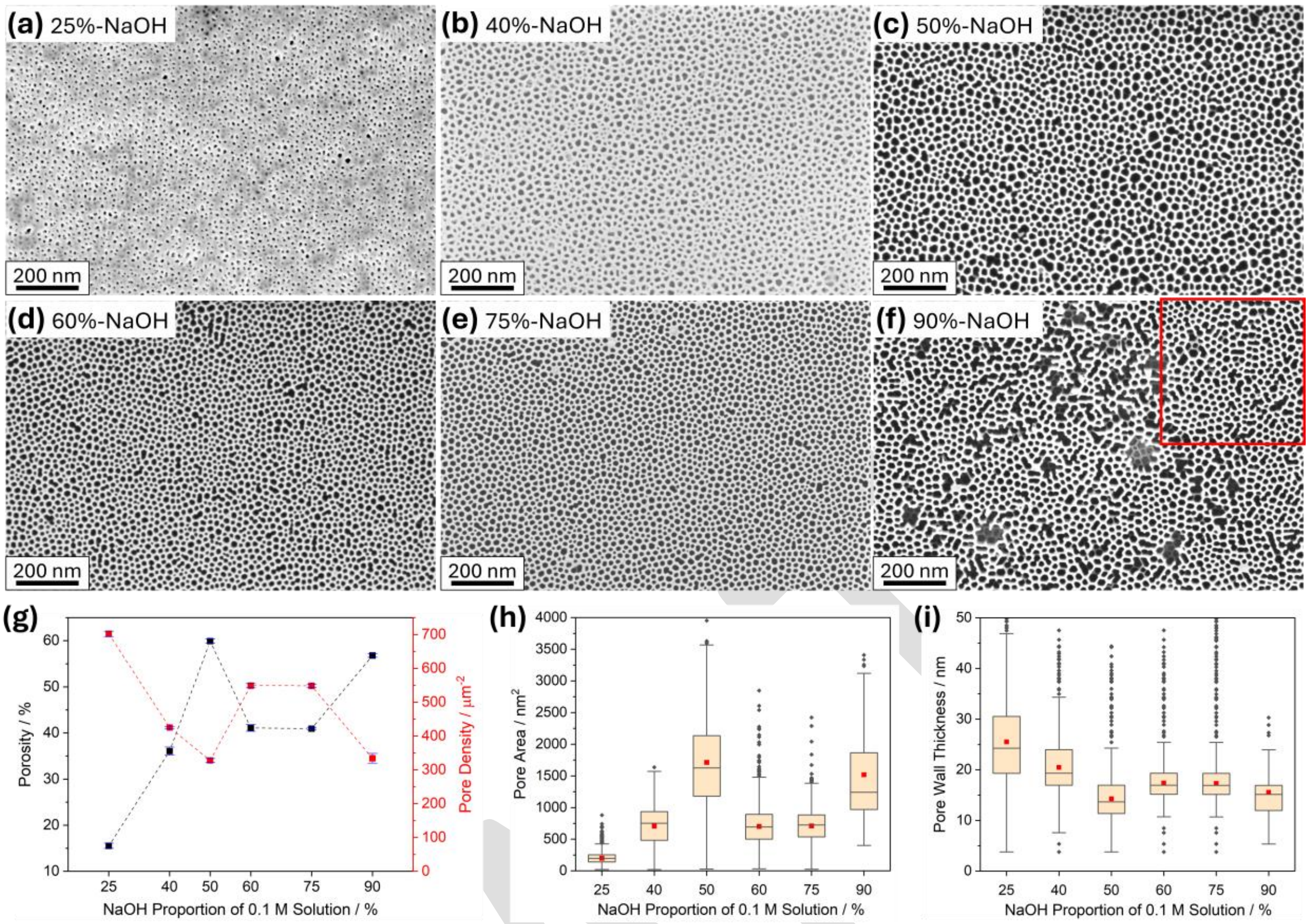

**Figure 3.** SE plan-view images, after removal of the uppermost 500 nm of the porous GaN layer by FIB milling, of 1 µm *n*-type GaN ([Si] = $1.5 \times 10^{19}$ $cm^{-3}$) electrochemically etched at 8 $V_{Ag/AgCl}$ in a mixture of (a) 0.075 M $Na_2C_2O_4$ and 0.025 M NaOH, (b) 0.06 M $Na_2C_2O_4$ and 0.04 M NaOH, (c) 0.05 $Na_2C_2O_4$ and 0.05 M NaOH, (d) 0.04 M $Na_2C_2O_4$ and 0.06 M NaOH, (e) 0.025 M $Na_2C_2O_4$ and 0.075 M NaOH, (f) 0.01 M $Na_2C_2O_4$ and 0.09 M NaOH. The images a-f have been used to obtain porous structure statistics for each of the ECE conditions: (g) porosity and mean pore density, error bars show the calculated standard error; (h) pore area, (i) pore wall thickness. Note that, due to potential FIB damage, only the area of image (f) within the annotated red box has been used to obtain porous structure statistics. For the box plots in (h) and (i), red squares represent the mean values and black lines within the upper and lower quartiles represent the median values. Note that some outliers at high values have been cropped from the plots.

## Columnar Pore Morphology Control: Identifying the Role of Hydroxide Ions

To further confirm whether the columnar pore morphology is controlled by NaOH, $Na_2C_2O_4$, or the specific combination of both etchants, mixed solutions of $OH^-$ with alternative etchants were investigated. Stronger bases, $Na_2CO_3$ and $Na_3PO_4$, were selected, eliminating the need for addition of NaOH to the mixture as their higher basicity results in greater intrinsic $OH^-$ concentrations relative to $Na_2C_2O_4$, due to anion hydrolysis. In fact, the earlier discussion on the irreproducibility of NaOH electrolytes, attributed to $CO_2$ absorption and the consequent formation of $CO_3^{2-}$, can be considered a similar mixed $OH^-$-$CO_3^{2-}$ system to $Na_2CO_3$ solutions, albeit with different relative compositions of each anion. $Na_2CO_3$ and $Na_3PO_4$ possess $pK_{b2}$ and $pK_{b3}$ values of 3.67 and 1.68,[26] respectively, resulting in 0.1 M solutions containing ~0.0045 M $OH^-$ (equivalent to a 4.5%-NaOH solution from the NaOH-$Na_2C_2O_4$ mixed solutions) and ~0.0364 M $OH^-$ (equivalent to 36.4%-NaOH), respectively (Equations 3 and 4).

$$CO_3^{2-} + H_2O \rightleftharpoons HCO_3^- + OH^- \quad (3)$$

$$PO_4^{3-} + H_2O \rightleftharpoons HPO_4^{2-} + OH^- \quad (4)$$

ECE performed using both $Na_2CO_3$ and $Na_3PO_4$ yields columnar pore morphologies similar to those resulting from the mixed NaOH-$Na_2C_2O_4$ solution (Figure 4). This suggests that the $OH^-$ anion plays a key role in directing the formation of vertically-aligned columnar pores, consistent with the vertical pore structures formed when NaOH is used alone. In addition, ECE in 0.1 M $Na_2CO_3$ solution yields the characteristic columnar pore morphology despite the low (~4.5%) $OH^-$ concentration, for which the equivalent $OH^-$ concentration in the NaOH-$Na_2C_2O_4$ mixture was too small. This, once again, indicates a significant difference in ECE activities between different anion species, and that the ECE mechanism resulting in the formation of columnar pores is dependent on the partner anion to $OH^-$ as well as requiring $OH^-$ presence, supporting the possibility of a two-step mechanistic reaction. Interestingly, the RMS roughness (measured over the 1 $\mu m^2$ AFM scan areas shown in Figure 4) of the resulting porous sample surface from ECE using $Na_2CO_3$ or $Na_3PO_4$ is notably higher than when using a mixed NaOH-$Na_2C_2O_4$ solution, with values of 13.8 and 11.7 nm for $Na_2CO_3$ and $Na_3PO_4$, respectively.

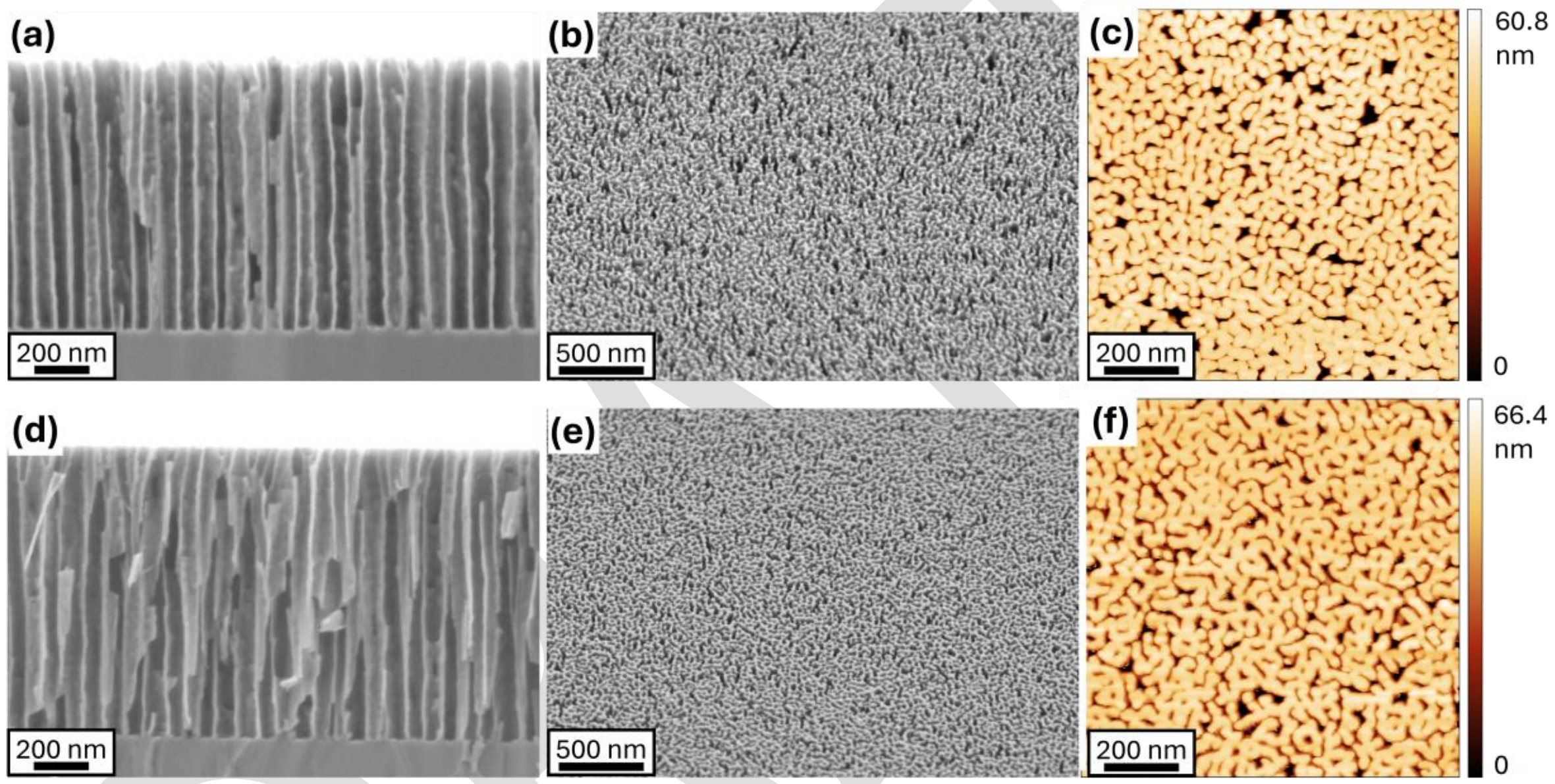

**Figure 4.** SE cross-sectional, SE plan-view, and AFM images of 1 µm *n*-type GaN ([Si] = $1.5 \times 10^{19}$ $cm^{-3}$) electrochemically etched at 8 $V_{Ag/AgCl}$ in (a–c) 0.1 M $Na_2CO_3$ and (d–f) 0.1 M $Na_3PO_4$. (a,d) SE cross-sectional images; (b,e) SE plan-view images acquired with the sample tilted by 45°; (c,f) AFM images.

Notably, the occurrence of a new mechanism upon mixing two etchants appears unique to $OH^-$-containing systems. Previous work from Harris-Lee *et al.*[14], and additional examples in Figure S4a,b, shows that alternative mixed etchant combinations (containing no or negligible $OH^-$ concentrations) did not yield fundamentally new pore morphologies, instead producing structures that are dominated by one etchant. Alternatively, Figure S4c shows that mixing $Na_2SO_4$ (an aggressive etchant that destroys the sample when used alone[14]) with NaOH produces the characteristic columnar morphology. This further supports the notion of a two-step ECE mechanism for all ECE conditions, where mixing two different anions only results in a new mechanism if different anions are favoured for different mechanistic steps; otherwise, the resulting pore morphology is simply that of the more dominant anion. So far, for aqueous electrolytes, it appears only mixtures containing $OH^-$ anions result in a new morphology, suggesting that the anion itself in single etchant systems (not including NaOH) is not involved in the first mechanistic step, regardless of $OH^-$ presence or not (i.e., no anion is required but $OH^-$ uniquely can be involved, or perhaps $H_2O$ will assume the role in the absence of $OH^-$).

**Columnar Pore Morphology Control: Influence of the Applied Potential**

So far, a bias potential of 8 $V_{Ag/AgCl}$ has been applied for all mixed etchant experiments. To examine the influence of applied potential, a series of bias potentials ranging from 4 to 10 $V_{Ag/AgCl}$ were employed for a 0.1 M $Na_2CO_3$ etchant solution (Figure 5). At 4 $V_{Ag/AgCl}$, the resulting pore morphology closely resembles that obtained from ECE in NaOH alone at the same bias potential (Figure 1a), suggesting that under these conditions only $OH^-$ anions are active, while the $CO_3^{2-}$ and $HCO_3^-$ anions (generated through $Na_2CO_3$ dissociation and hydrolysis) remain relatively inactive. Interestingly, although areas lacking visible porosity are present in both cases, the surface pores produced in $Na_2CO_3$ at 4 $V_{Ag/AgCl}$ appear less hexagonal and faceted than those formed in NaOH alone (Figure 1d).

As the applied potential is increased to 5 and 6 $V_{Ag/AgCl}$, the vertically-aligned pores widen and the surface evolves toward a more uniformly porous morphology, possessing smaller individual pores with greater density and relatively uniform spatial distribution. Increasing applied potential to the 7-9 $V_{Ag/AgCl}$ range, the pore morphology becomes increasingly columnar, with individual columns continuing to widen as the applied potential is increased. In addition, the surface morphology coarsens and becomes progressively more interconnected, despite the columnar pores within the bulk appearing to remain spatially isolated, exhibiting no overlap or interconnection. This control over pore properties is critical for the application of porous GaN across a range of technologies, for example, in infiltration processes requiring open surface access and easy-to-penetrate bulk pores, or in strain relaxation strategies where a closed surface with a highly porous subsurface may be preferable for both overgrowth and structure flexibility.[4] While the unusual surface morphology observed at 4 $V_{Ag/AgCl}$ resulted in an RMS roughness (measured over the 1 $\mu m^2$ AFM scan areas shown in Figure 5) of 6.0 nm, the remaining samples exhibited an increase in RMS roughness with increasing applied potential, from 0.8 and 0.9 nm at 5 and 6 $V_{Ag/AgCl}$, respectively, to 4.1, 13.8, and 16.6 nm for 7, 8, and 9 $V_{Ag/AgCl}$, respectively.

At 10 $V_{Ag/AgCl}$, the pore morphology becomes non-uniform and the porosified layer appears to crack or partially delaminate, preventing further porosification (Figure S5a,b). The surface is highly roughened on the microscale and appears to have been partially removed, as indicated by the exposure of the bulk columns rather than a surface morphology comparable to those observed at lower applied potential.

Counterintuitively, chronoamperometry data reveals that increasing the applied potential results in a longer ECE duration (Figure S5c). Increased bias should result in a greater number of generated holes, less recombination, and an accelerated electrochemical reaction. The increase in ECE duration with increasing applied potential may therefore indicate a change in the relative proportions of competing ECE mechanisms, from the rapid ECE observed in NaOH-dominated conditions towards the slower regime associated with the mixed-electrolyte composition. For example, a combination of the $OH^-$ alone, the $CO_3^{2-}$ alone, and the $OH^-$-$CO_3^{2-}$ mixed ECE mechanisms may be occurring simultaneously, but as the applied potential is increased, different mechanisms are favoured and the relative contribution of each of these changes. In addition to the surface chemistry affecting the resulting pore morphology, previous work by Chen *et al.* proposed that linear pores result under conditions in which the ratio of the inter-pore spacing ($d_w$) and the GaN space charge region ($d_{Sc}$) are less than one (i.e., $d_w/d_{sc} < 1$).[29] In this case, increasing the applied potential would increase $d_{sc}$, making the pores more linear and less branched, hence matching the trend shown in Figure 5.

However, the $d_w/d_{sc}$ ratio alone does not account for the chronoamperometry data or several other empirical observations. For example, ECE with 0.1 M $Na_2C_2O_4$ at 8 $V_{Ag/AgCl}$ ([Si] = $1.5 \times 10^{19}$ $cm^{-3}$) results in substantial GaN removal, a highly roughened surface, and a highly porous structure of very poor uniformity. Further, when mixed with NaOH, the resulting pore structure then becomes well defined despite there being no change in applied potential (Figure 1), hence a change in $d_{sc}$ related to a change in potential cannot be the explanation in this case. These observations provide significant evidence towards the influence of surface chemistry, and are consistent with a shift in the dominant ECE mechanism with increasing applied potential; although, importantly, they do not preclude the influence of the $d_w/d_{sc}$ ratio. Furthermore, changes in the applied potential may alter the interactions between the electrolyte anions and the GaN surface, thereby modifying the surface chemistry and band bending, consequently affecting the $d_w/d_{sc}$ ratio beyond the direct effect of applied potential on the space-charge region. It is therefore likely that both surface chemistry and the local electric field contribute to the resulting pore morphology from ECE.

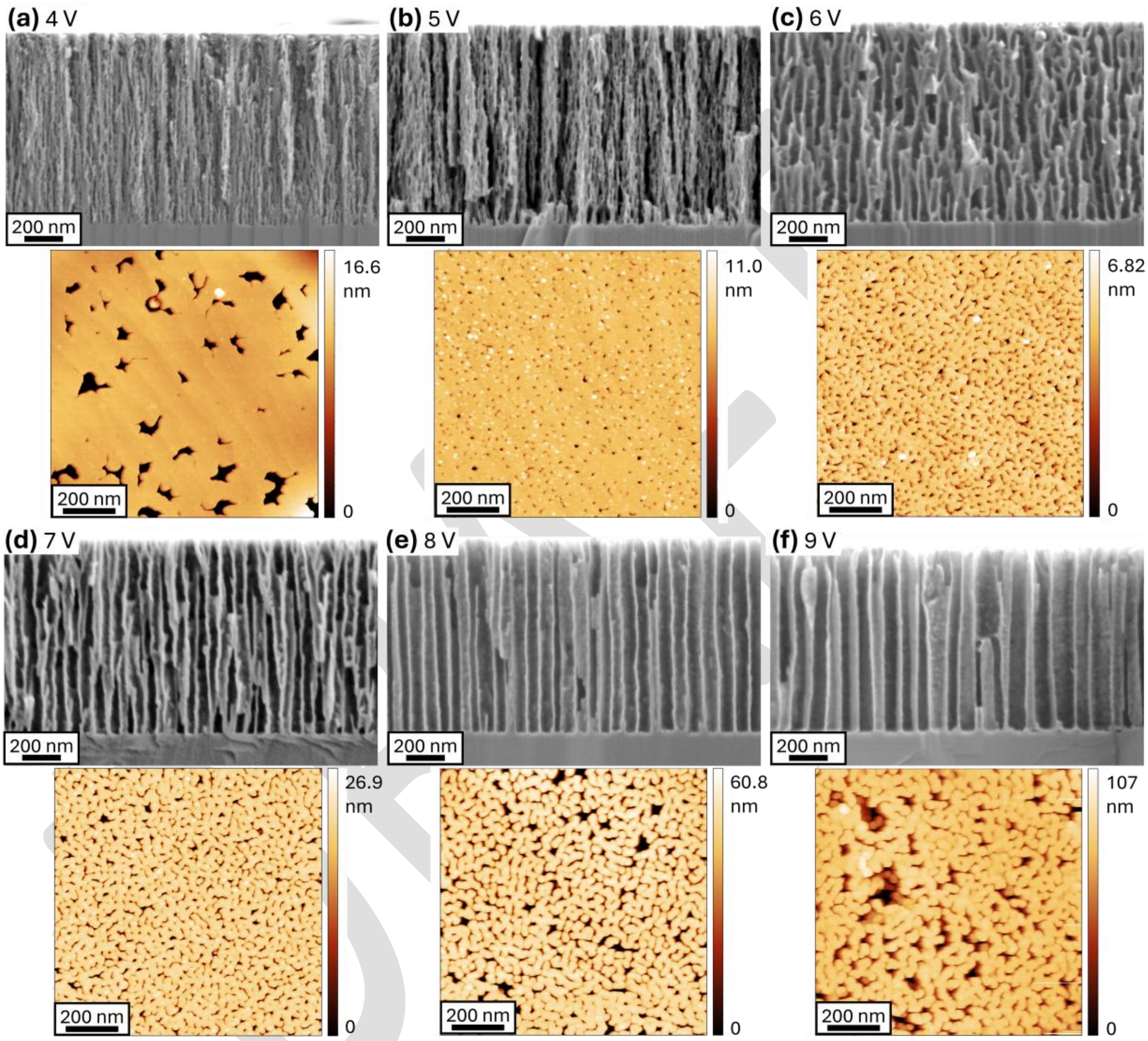

**Figure 5.** (top) SE cross-section images, (bottom) AFM surface images for 1 µm *n*-type GaN ([Si] = $1.5 \times 10^{19}$ $cm^{-3}$) electrochemically etched in 0.1 M $Na_2CO_3$ at (a) 4 $V_{Ag/AgCl}$, (b) 5 $V_{Ag/AgCl}$, (c) 6 $V_{Ag/AgCl}$, (d) 7 $V_{Ag/AgCl}$, (e) 8 $V_{Ag/AgCl}$, (f) 9 $V_{Ag/AgCl}$.

To examine the variations in pore morphology from different ECE bias potentials in greater detail, 3D renders of single pores have been reconstructed from FIB-SEM tomographs for porous structures obtained from ECE with 0.1 M $Na_2CO_3$ electrolyte at constant applied potentials of 4, 6, and 8 $V_{Ag/AgCl}$ (**Figure 6**a,b,c, respectively). The full tomographic datasets can be found in the Supporting Information as Movie S1, Movie S2, and Movie S3, respectively. For samples electrochemically etched with a bias potential of 4 $V_{Ag/AgCl}$, each surface pore appears to rapidly expand laterally to fill space below the significant amount of unetched area across the surface. This is similar to that seen previously by Zhang *et al.*[21] for ECE of *n*-type GaN in 0.25 M oxalic acid, where certain applied voltage/doping density regimes resulted in significant pore lateral spread beneath a relatively low density of surface pits. Porous GaN-based distributed Bragg reflector structures, electrochemically etched using top-down ECE protocols, also show a similar phenomenon, where a dislocation is etched in the NID layer followed by rapid lateral expansion through the doped layer.[28,30] Each laterally expanding branch is only terminated by meeting another branch originating from a different surface pore (i.e., the pores fill the available space, but do not overlap and interconnect). Expansion also occurs vertically at a seemingly similar rate, hence when lateral expansion finishes, each branch only then extends vertically, resulting in a highly vertically-oriented bulk pore structure. For pores fabricated using a 6 $V_{Ag/AgCl}$ bias potential, a similar effect is seen, but with the initial surface pores much smaller and denser, resulting in less room to expand laterally subsurface and hence fewer vertical pores forming from a single surface pore, albeit with increased pore diameter.

Each pore formed using 4 $V_{Ag/AgCl}$ and 6 $V_{Ag/AgCl}$ bias potentials is spatially isolated from all other pores throughout the entire porous layer (Figure 5a,c), hence the 3D models provided are for a single isolated pore from the structure. In contrast, pores formed using 8 $V_{Ag/AgCl}$ bias potential are all interconnected at the surface (Figure 5e), hence applying this same method would result in the entire porous structure being chosen. However, each surface pore does not branch laterally to fill the space below unetched surface, like samples etched at 4 $V_{Ag/AgCl}$ and 6 $V_{Ag/AgCl}$ do. There is therefore a more immediate initiation of each columnar pore, and hence the 3D pore model has been restricted to only show one of these columns. Uniform, relatively wide columnar morphologies, like those seen in **Figure 6**c, will improve electrolyte and ECE product diffusion within the pores compared to thinner pores with lots of branching, making these structures promising for applications that are significantly restricted by diffusion effects (e.g., catalysis, sensing),[31,32] as well as for impregnation with alternative materials to form composites.[10,11,33]

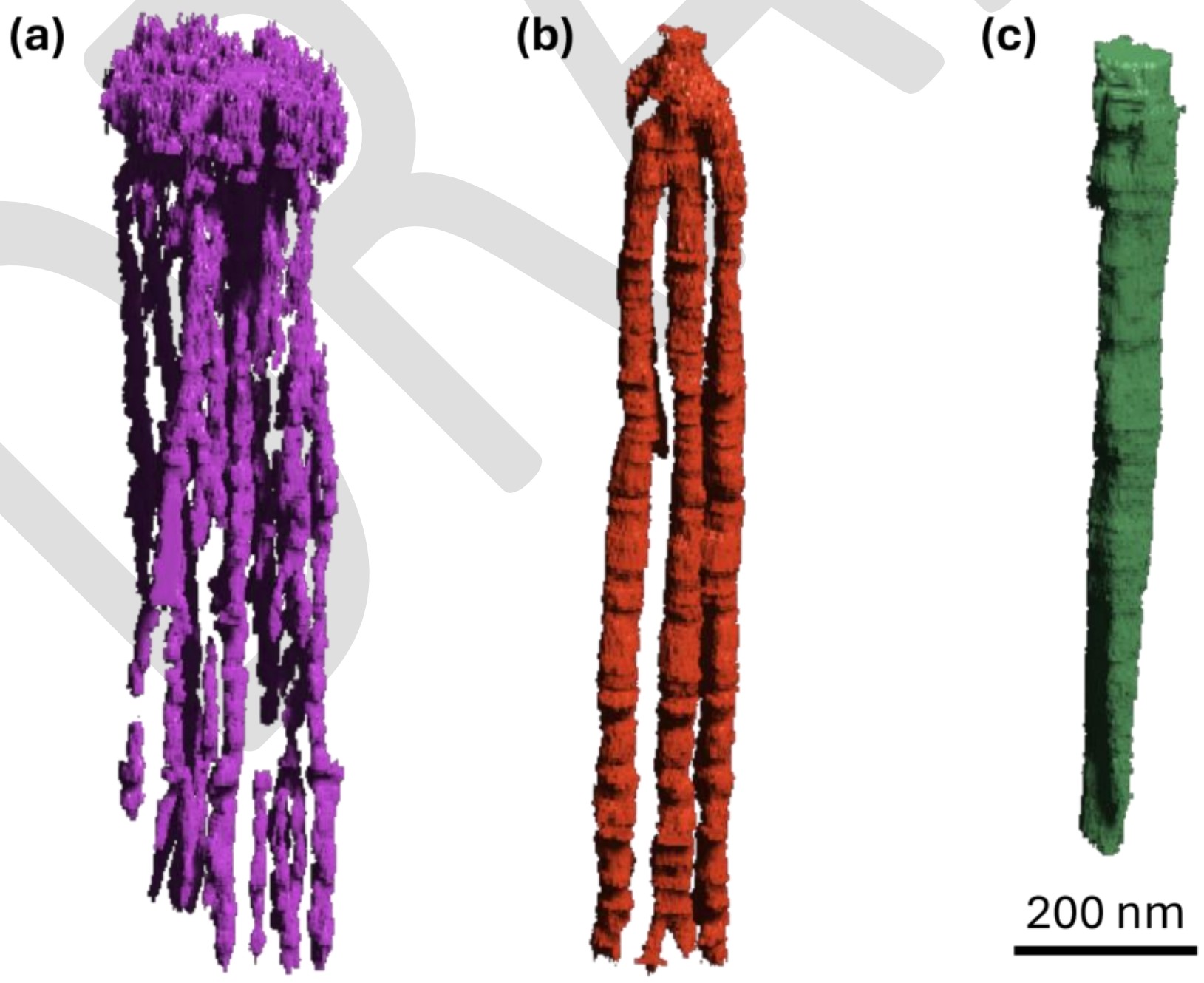

**Figure 6.** 3D renders of single, spatially isolated pores from the ECE of 1 µm *n*-type GaN ([Si] = $1.5 \times 10^{19}$ $cm^{-3}$) in 0.1 M $Na_2CO_3$ at (a) 4 $V_{Ag/AgCl}$, (b) 6 $V_{Ag/AgCl}$, (c) 8 $V_{Ag/AgCl}$, reconstructed from FIB-SEM tomographic datasets which can be found in the Supporting Information as Movie S1, Movie S2, and Movie S3, respectively.

The strong potential dependence of ECE in mixed etchant solutions can be beneficially exploited by stepping the applied potential during a single ECE process, resulting in a porous GaN layer composed of distinct, stepped porous layers that directly correspond to each applied potential (Figure 7). This is analogous to work previously carried out by Griffin and Oliver on voltage-stepped ECE in oxalic acid.[12] When starting at high bias potential (8 $V_{Ag/AgCl}$), each of the three voltage regimes can be distinguished within the porous structure (Figure 7a,b, red lines mark different bias potential regimes), and the surface appears qualitatively the same as that for a full ECE process at 8 $V_{Ag/AgCl}$ (Figure 5e, Figure S5h). Interestingly, this is not the case when starting the stepped ECE protocol at the low bias potential (4 $V_{Ag/AgCl}$). Figure 7c does not contain the same distinct steps in morphology between pore structures obtained with 4, 6, and 8 $V_{Ag/AgCl}$ potentials alone, particularly the structure resulting from a 6 $V_{Ag/AgCl}$ bias potential cannot be clearly distinguished. Further, the surface does not appear identical to that from ECE under 4 $V_{Ag/AgCl}$ only (Figure 5a, Figure S5d). However, the structure in Figure 7c is promising proof of concept for creating ordered, highly porous subsurface architectures while retaining a significantly smoother surface, far more suitable for overgrowth. It is unclear why starting at the highest bias potential results in a more distinct stepped morphology, and particularly, if increasing the potential to 8 $V_{Ag/AgCl}$ results in re-etching the 6 $V_{Ag/AgCl}$ region to look like that from the higher potential, why this does not also occur for the 4 $V_{Ag/AgCl}$ region. However, overall, this preliminary data shows that stepping the voltage throughout the ECE process can be used to create unique layered structures, controlling the bulk and the surface morphology throughout the ECE process. This highlights the promise for more advanced ECE strategies based on mixed etchant solutions, in which varying applied potential can alter the dominant ECE mechanism and hence the resulting pore morphology. Such approaches offer enhanced control over porous layer and surface architectures and enable the fabrication of more complex, spatially structured porous GaN systems; critical techniques for tailoring and advancing porous GaN application.

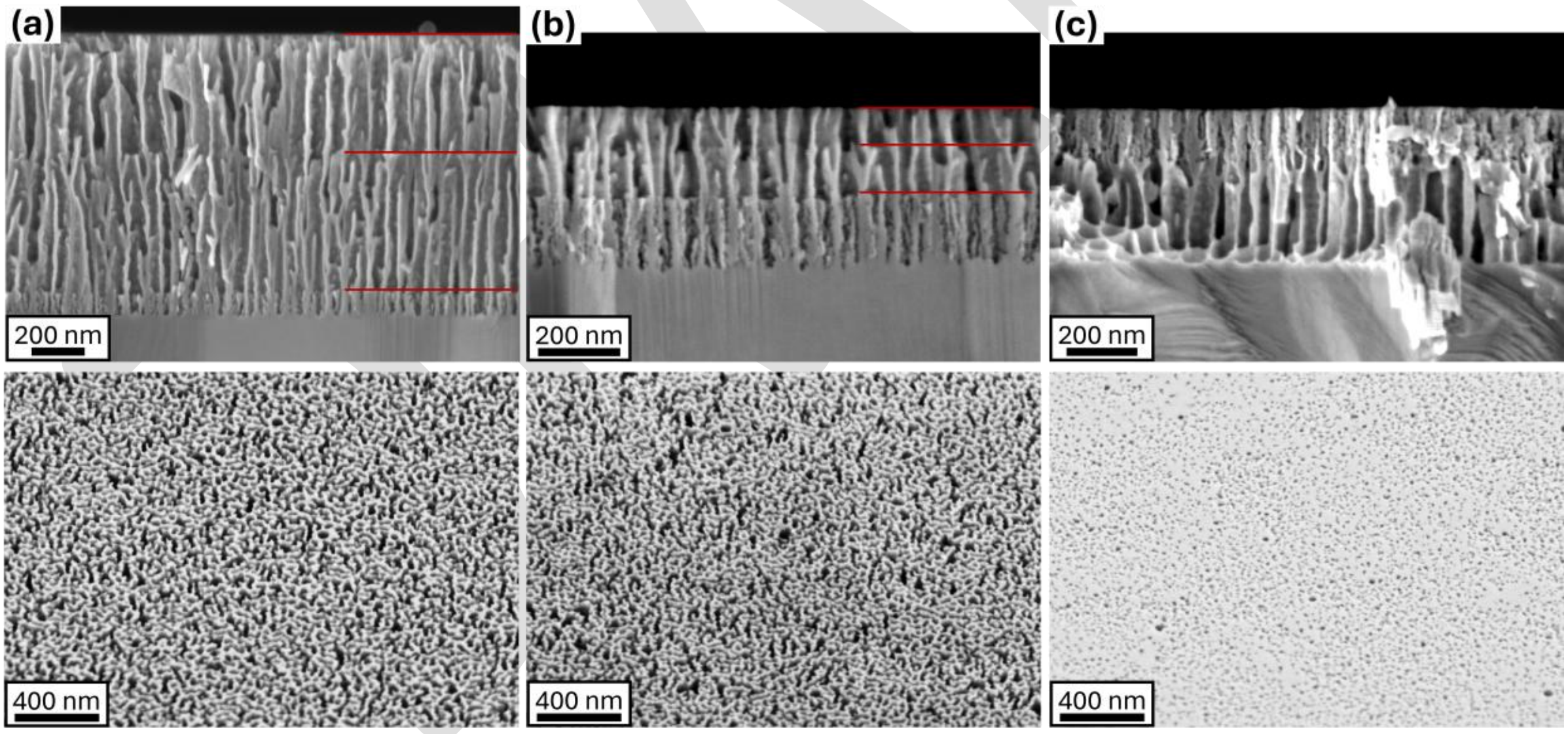

**Figure 7.** SE images (top) cross-section, (bottom) 45° tilted plan-view for 1 µm *n*-type GaN ([Si] = $1.5 \times 10^{19}$ $cm^{-3}$) electrochemically etched in 0.1 M $Na_2CO_3$ using stepped potentials of (a) 8 $V_{Ag/AgCl}$ for 100 s, then 6 $V_{Ag/AgCl}$ for 100 s, then 4 $V_{Ag/AgCl}$ for 100 s; (b) 8 $V_{Ag/AgCl}$ for 30 s, then 6 $V_{Ag/AgCl}$ for 30 s, then 4 $V_{Ag/AgCl}$ for 100 s; (c) 4 $V_{Ag/AgCl}$ for 30 s, then 6 $V_{Ag/AgCl}$ for 30 s, then 8 $V_{Ag/AgCl}$ for 100 s. Red lines show distinct pore morphology changes as a result of changing the bias potential.

## Conclusions

The incorporation of $OH^-$ anions, either by addition of NaOH or by hydrolysis in water, into non-acidic etchant solutions directs the ECE of *n*-type *c*-plane GaN toward vertically-oriented, columnar pore morphologies. Using mixtures of NaOH with $Na_2C_2O_4$, the ECE process appears to be controlled by a new mechanism, rather than by the simultaneous mechanisms of NaOH and $Na_2C_2O_4$ alone. Basic electrolytes such as $Na_2CO_3$ and $Na_3PO_4$, which intrinsically generate significant concentrations of $OH^-$ through hydrolysis, also yield columnar pore morphologies when used alone (i.e., without deliberate addition of $OH^-$ anions, for example from NaOH in this work, into the mixture). From these results, an overall mechanism for ECE of *n*-type GaN, which is currently unknown and poses a significant hindrance to the advancement of porous GaN research and development, was derived as a multi-step process where each of the $OH^-$ and $C_2O_4^{2-}$ (or $CO_3^{2-}$, $PO_4^{3-}$) anions are favoured for different steps. Further, due to the increased ligating strength of $C_2O_4^{2-}$ compared to $OH^-$ towards $Ga^{3+}$ cations, it is inferred that the first step must be controlled by the $OH^-$ anion, forming a reactive intermediate, followed by a second step that involves the complete removal of $Ga^{3+}$ from the lattice by $C_2O_4^{2-}$. This is consistent with empirical trends seen in previous works when utilising a range of different electrolytes, and enables electrolytes to be categorised into groups based on the resulting pore morphology, which is consistent with the electrolytes' properties. This grouping is a significant step in the advancement of understanding and predicting the ECE mechanism.

The columnar pore widths, separation, and total porous layer porosity can be readily tuned to suit specific application requirements by adjusting the ratio of NaOH within the mixed etchant composition. Variation of the applied bias potential also enables systematic control over pore morphology, ranging from narrow, vertically-aligned pores characteristic of NaOH alone, to wider columnar structures. This allows dynamic, *in situ* modulation of pore architecture during a single ECE process to create multi-layered porous structures or structures with graded pore sizes.

The resulting columnar porous GaN architectures are particularly well suited for applications including composite formation *via* pore infiltration, devices requiring high porosity (e.g., for electrical or thermal resistance), strain relaxation layers for epitaxial overgrowth (particularly when controlling surface morphology by modulating bias potential), and ordered, high porosity architectures for piezoelectric applications. Overall, this work establishes a tuneable ECE framework that enables precise and scalable control over porous GaN architectures, significantly advancing the design space for porous GaN-containing devices.

## Methods

**GaN Sample Preparation**

Ga-polar *c*-plane GaN epilayer growth was performed by metal–organic vapor phase epitaxy (MOVPE) in an Aixtron close-coupled showerhead reactor on 2-inch (0001) sapphire substrates. The sample structure consisted of a 2.4 µm-thick layer of nominally undoped GaN, followed by a 1 µm-thick layer of highly *n*-type GaN with a Si concentration ([Si]) of $1.5 \times 10^{19}$ $cm^{-3}$. The threading dislocation density was measured at ca. $6 \times 10^{8}$ $cm^{-2}$ using the method described by Oliver et al.[34]

**Chemical Reagents and Electrode Materials**

Sodium hydroxide (NaOH, Thermo Fisher Scientific, >98%), sodium oxalate ($Na_2C_2O_4$, Thermo Fisher Scientific, 99%), sodium carbonate ($Na_2CO_3$, Alfa Aesar, 98%), sodium phosphate tribasic ($Na_3PO_4$, Sigma-Aldrich, 96%), sodium sulphate ($Na_2SO_4$, Sigma-Aldrich, >99%), were used as supplied by the manufacturer. All aqueous solutions were prepared with ultrapure deionized (DI) water (resistivity = 22.2 MΩ cm at 25 °C) from a Direct-Q 3 UV water purification system (Milli-Q).

A (10 × 10 × 0.1) $mm^3$ platinum plate (Ossila, 99.99%) was used as the counter electrode (CE). A 3 M Ag/AgCl electrode (Ossila) was used as the reference electrode (RE). Note that, unless otherwise stated, all potentials herein are referenced to the 3 M Ag/AgCl RE scale. Indium wire (In, Thermo Fisher Scientific, >99.998%) was melted and soldered to the *n*-type GaN surface to form the conductive contact.

**Electrochemical Etching**

ECE was carried out on a Gamry Reference 3000 potentiostat connected to a three-electrode electrochemical cell containing a Pt plate CE and 3 M Ag/AgCl RE. Electrical connection was formed on an ∼(0.6 × 1.2) $cm^2$ GaN sample by soldering an indium metal contact on one end, which was then connected into the electrochemical cell as the working electrode (WE) and submerged into electrolyte approximately halfway. 50 mL of fresh 0.1 mol $dm^{-3}$ (M) solutions were used for every experiment, unless stated otherwise. For mixed electrolyte solutions, the total solution concentration was 0.1 M, and percentage compositions of each electrolyte refer to the molar percent of each electrolyte added to solution initially. During ECE, etchant solutions were stirred at 200 rpm using a magnetic stirrer bar. After ECE, samples were washed thoroughly with DI water, followed by drying in $N_2$.

Chronoamperometry curves display the current generated during the ECE at a constant applied voltage as a function of time. Herein, for accurate comparison of etching current magnitudes between samples, all etching currents ($I$) have been normalized to the original planar geometric surface area of the etched sample region. Note that, due to the ongoing porosification process, the sample surface area is continuously increasing; hence, the measured current cannot be meaningfully converted to current density.

**Physical Characterisation**

Cross-sectional and surface pore morphologies were obtained by scanning electron microscopy (SEM, Zeiss Gemini 300) using secondary electron (SE) imaging. Unless otherwise stated, all plan-view surface images were taken at a 45° sample tilt. Surface topography images were generated by atomic force microscopy (AFM, Bruker Dimension Icon) in PeakForce tapping mode by using ScanAsyst HPI tips from Bruker.

FIB-SEM tomography was performed using a Zeiss Crossbeam 540 with a Zeiss Gemini 2 column applied in a plan-view geometry, as described comprehensively by Thornley *et al*.[35] In brief, each SE image used to produce the tomograph is taken in a plan-view orientation, rather than the more conventional cross-section orientation. This ensures that the longest pore dimension within the vertically-oriented columns is perpendicular to the imaged plane, therefore limiting issues relating to the pore back effect.[36] The final tomographic datasets for plan-view (input frames) and cross-section (reconstructed images) are presented as Movie S1, Movie S2, and Movie S3 in the Supporting Information, for samples fabricated with ECE conditions of 4, 6, and 8 $V_{Ag/AgCl}$ bias potential, respectively, with 0.1 M $Na_2CO_3$ electrolyte.

Plan-view SE images of the bulk (i.e., subsurface) pore morphologies were recorded by removing the uppermost 500 nm of the 1 μm-thick porous GaN layer using an analogous FIB arrangement as that used by Thornley *et al.* for plan-view tomography.[35] Plan-view SE images were then recorded separately, perpendicular to the new 'surface'. Performing milling and imaging separately also ensured that no tilt would be present in either the milled plane or the imaging plane. Key properties for each porous GaN sample (porosity, pore density, pore area, pore wall thickness) were then determined from these images.

Porosity calculations were performed by converting each analysed plan-view SE image into a binary image using the threshold function in ImageJ software such that the pore (etched) and pore wall (unetched) regions were separated. Each ∼(2844 × 1922) $nm^2$ binarised SE image was then divided into nine equal (948 × 640.7 $nm^2$) regions (SI, Figure S1a). The porosity (etched area/total area) for each of the nine regions was then calculated, and mean values were calculated. Standard errors were determined and used for error bars. Pore density calculations used the same nine regions, counting the number of pores in each area to calculate the mean density and standard error (used as error bars). Pores overlapping different regions were only counted in the region where they possessed the highest total area (SI, Figure S1a; white colouring shows where a pore has not been counted). Pores overlapping with the border of the entire ∼(2844 × 1922) $nm^2$ SE image were only counted if they had a visible area greater than half the mean pore area of the entire image (SI, Figure S1a; white colouring shows where a pore has not been counted).

Mean pore areas were calculated by binarising each plan-view SE image, measuring the total number of pixels of each pore, then converting to $nm^2$ using the original image scale. Due to the high number of pore samples, standard errors were extremely small. Areas were used instead of diameters due to the non-circular shape of the pores' cross-sections preventing accurate conversion.

Pore wall thickness distributions were determined by skeletonization of the wall network combined with a Euclidean distance transform of a binarised SE image. In detail, the pore and pore wall regions were first defined by image binarization (white pixels representing the pore walls, black pixels representing the pores). The pore wall region 'skeleton' was then found by iterative morphological thinning, which repeatedly peels away the outer boundary pixels until it is reduced to a one-pixel-wide medial axis, containing all pore wall pixels that have more than one closest point to the pore wall boundary (i.e., to the pores). A Euclidian distance transform was then applied, assigning each pore wall pixel its distance, in pixels, to the nearest pore pixel. The distance assigned to each skeleton pixel therefore corresponds to the distance from the pore wall centreline (the skeleton) to the nearest pore. The local pore wall thickness for each skeleton pixel was obtained by multiplying this distance by two, since, by definition, the skeleton pixel must be equidistant from more than one closest pore pixel. The resulting values were then converted from pixels to nanometres using the known scale of the original image. This procedure was applied to every skeleton pixel, generating a large dataset of local pore wall thickness values across the entire analysed image.

## Supporting Information

Pore and pore wall identification and characterisation for statistical analysis of different pore morphologies (Figure S1); SEM and chronoamperometry characterisation when using NaOH and $Na_2C_2O_4$ at different applied potentials and solution ages (Figure S2); SEM and chronoamperometry characterisation after ECE in varying relative ratios of mixed NaOH-$Na_2C_2O_4$ electrolyte (Figure S3); SEM characterisation after ECE in different mixed electrolytes (Figure S4); SEM and chronoamperometry characterisation after ECE in $Na_2CO_3$ with varying applied potential (Figure S5); tomographic datasets for plan-view (input frames) and cross-section (reconstructed images) for samples fabricated with ECE conditions of 4 $V_{Ag/AgCl}$ (Movie S1), 6 $V_{Ag/AgCl}$ (Movie S2), and 8 $V_{Ag/AgCl}$ (Movie S3) bias potential, with 0.1 M $Na_2CO_3$ electrolyte.

## Acknowledgments

This research was supported by the Royal Academy of Engineering under the Chairs in Emerging Technologies Scheme, which is sponsored by the Department for Science, Innovation and Technology (DSIT). Funding was also received from the EPSRC under EP/X015300/1. BT would like to acknowledge the Ernest Oppenheimer Trust at the University of Cambridge and JZ would like to acknowledge support from the Ministry of Business, Innovation and Employment (MBIE), New Zealand, Strategic Science Investment Fund 'Advanced Energy Technology Platforms' under contract No. RTVU2004. Atomic force microscopy studies employed equipment supported by the Cambridge Royce facilities grant EP/P024947/1 and the Sir Henry Royce Institute - recurrent grant EP/R00661X/1. We acknowledge the support of the Wolfson Electron Microscopy Suite and the use of the Zeiss Crossbeam 540 funded by Royce under grant EP/R008779/1.

## References

[1] C.J. Humphreys, "Solid-State Lighting," MRS Bull. **33**(4), 459–470 (2008).

[2] H. Amano, Y. Baines, E. Beam, M. Borga, T. Bouchet, P.R. Chalker, M. Charles, K.J. Chen, N. Chowdhury, R. Chu, C. De Santi, M.M. De Souza, S. Decoutere, L. Di Cioccio, B. Eckardt, T. Egawa, P. Fay, J.J. Freedsman, L. Guido, O. Häberlen, G. Haynes, T. Heckel, D. Hemakumara, P. Houston, J. Hu, M. Hua, Q. Huang, A. Huang, S. Jiang, H. Kawai, D. Kinzer, M. Kuball, A. Kumar, K.B. Lee, X. Li, D. Marcon, M. März, R. McCarthy, G. Meneghesso, M. Meneghini, E. Morvan, A. Nakajima, E.M.S. Narayanan, S. Oliver, T. Palacios, D. Piedra, M. Plissonnier, R. Reddy, M. Sun, I. Thayne, A. Torres, N. Trivellin, V. Unni, M.J. Uren, M. Van Hove, D.J. Wallis, J. Wang, J. Xie, S. Yagi, S. Yang, C. Youtsey, R. Yu, E. Zanoni, S. Zeltner, and Y. Zhang, "The 2018 GaN power electronics roadmap," J. Phys. D Appl. Phys. **51**(16), 163001 (2018).

[3] R.A. Oliver, "Critical Assessment 23: Gallium nitride-based visible light-emitting diodes," Materials Science and Technology **32**(8), 737–745 (2016).

[4] Y. Ji, M. Frentrup, X. Zhang, J. Pongrácz, S.M. Fairclough, Y. Liu, T. Zhu, and R.A. Oliver, "Porous pseudo-substrates for InGaN quantum well growth: Morphology, structure, and strain relaxation," J. Appl. Phys. **134**(14), 145102 (2023).

[5] S.S. Pasayat, C. Gupta, Y. Wang, S.P. DenBaars, S. Nakamura, S. Keller, and U.K. Mishra, "Compliant micron-sized patterned ingan pseudo-substrates utilizing porous GaN," Materials **13**(1), (2020).

[6] D. Cao, H. Xiao, J. Fang, J. Liu, Q. Gao, X. Liu, and J. Ma, "Photoelectrochemical water splitting on Nanoporous GaN thin films for energy conversion under visible light," Mater. Res. Express **4**(1), (2017).

[7] D. Cao, H. Xiao, H. Xu, J. Cui, Q. Gao, and H. Pei, "Enhancing the photocatalytic activity of GaN by electrochemical etching," Mater. Res. Bull. **70**, 881–886 (2015).

[8] R. Radzali, Z. Hassan, N. Zainal, and F.K. Yam, "Preparation of porous InAlGaN/Si(111) by photoelectrochemical etching for high performance hydrogen gas sensors at room temperature," Sens. Actuators B Chem. **213**, 276–284 (2015).

[9] M.R. Zhang, and G.B. Pan, "Porous GaN electrode for anodic stripping voltammetry of silver(I)," Talanta **165**, 540–544 (2017).

[10] K.T.P. Lim, C. Deakin, B. Ding, X. Bai, P. Griffin, T. Zhu, R.A. Oliver, and D. Credgington, "Encapsulation of methylammonium lead bromide perovskite in nanoporous GaN," APL Mater. **7**(2), (2019).

[11] M. Zhang, Y. Liu, J. Wang, and J. Tang, "Photodeposition of palladium nanoparticles on a porous gallium nitride electrode for nonenzymatic electrochemical sensing of glucose," Microchimica Acta **186**(2), (2019).

[12] P.H. Griffin, and R.A. Oliver, "Porous nitride semiconductors reviewed," J. Phys. D Appl. Phys. **53**(38), 383002 (2020).

[13] A. Shushanian, D. Iida, Z. Zhuang, Y. Han, and K. Ohkawa, "Analysis of the n-GaN electrochemical etching process and its mechanism in oxalic acid," RSC Adv. **12**(8), 4648–4655 (2022).

[14] T.R. Harris-Lee, B. Thornley, J. Zhang, M.J. Kappers, and R.A. Oliver, "Porous GaN: Anion-Specific Electrochemical Etching Mechanisms and Morphological Control," ACS Appl. Mater. Interfaces **17**(47), 64931–64941 (2025).

[15] J. Liu, J. Cui, and H. Xiao, "Electrochemical etching of n-type GaN in different electrolytes," J. Alloys Compd. **983**, 173846 (2024).

[16] C. Zhang, G. Yuan, A. Bruch, K. Xiong, H.X. Tang, and J. Han, "Toward Quantitative Electrochemical Nanomachining of III-Nitrides," J. Electrochem. Soc. **165**(10), E513–E520 (2018).

[17] I. Medjahed, C. Licitra, S. Sadki, and C. Pernel, "Electrochemical Prediction Tool of Porous GaN Morphology," Journal of Physical Chemistry C **128**(15), 6249–6258 (2024).

[18] W.T. Huang, L.X. Hong, and R.S. Liu, "Nanostructure Control of GaN by Electrochemical Etching for Enhanced Perovskite Quantum Dot LED Backlighting," ACS Appl. Mater. Interfaces **15**(33), 39505–39512 (2023).

[19] A. Shushanian, D. Iida, Y. Han, and K. Ohkawa, "Chemical and morphological characterization of the anodic oxidation of *n* -GaN in inorganic electrolytes," New Journal of Chemistry **46**(48), 23013–23018 (2022).

[20] W.F. Lim, Z. Hassan, N.M. Ahmed, and H.J. Quah, "Porous Formation in p-Type Gallium Nitride Films via 50 Hz Operated Alternating Current-Assisted Photo-Electrochemical Etching in Methanol-Sulfuric Acid Solution," J. Electrochem. Soc. **165**(10), H620–H628 (2018).

[21] J. Zhang, B. Thornley, T.R. Harris-Lee, S.A. Dar, M.J. Kappers, and R.A. Oliver, "The impact of dislocation-mediated etching on the structure of porous GaN," J. Appl. Phys. **139**(17), (2026).

[22] M.J. Schwab, J. Han, and L.D. Pfefferle, "Neutral anodic etching of GaN for vertical or crystallographic alignment," Appl. Phys. Lett. **106**(24), (2015).

[23] W.W. Rudolph, and C.C. Pye, "Gallium(III) hydration in aqueous solution of perchlorate, nitrate and sulfate. Raman and 71-Ga NMR spectroscopic studies and ab initio molecular orbital calculations of gallium(III) water clusters," Physical Chemistry Chemical Physics **4**(18), 4319–4327 (2002).

[24] Ralph G. Pearson, "Hard and Soft Acids and Bases," J. Am. Chem. Soc. **85**(22), 3533–3539 (1963).

[25] S.A. Wood, and I.M. Samson, "The aqueous geochemistry of gallium, germanium, indium and scandium," Ore Geol. Rev. **28**(1), 57–102 (2006).

[26] S. Iida, H. Terabe, T. Tachibana, K. Wada, I. Mochizuki, A. Yagishita, T. Hyodo, and Y. Nagashima, *CRC Handbook of Chemistry and Physics*, 84th Edition (CRC Press, 2014).

[27] W.J. Tseng, D.H. Van Dorp, R.R. Lieten, P.M. Vereecken, and G. Borghs, "Anodic etching of n-GaN epilayer into porous GaN and its photoelectrochemical properties," Journal of Physical Chemistry C **118**(51), 29492–29498 (2014).

[28] B. Thornley, M. Sarkar, S. Ghosh, M. Frentrup, M.J. Kappers, T.R. Harris-Lee, and R.A. Oliver, "A cascade model for the defect-driven etching of porous GaN distributed Bragg reflectors," Acta Mater. **308**, 121957 (2026).

[29] D. Chen, H. Xiao, and J. Han, "Nanopores in GaN by electrochemical anodization in hydrofluoric acid: Formation and mechanism," J. Appl. Phys. **112**(6), (2012).

[30] T.R. Harris-Lee, Y. Zhang, B. Thornley, J. Zhang, M.J. Kappers, and R.A. Oliver, "Analysis of in situ electrochemical characterization methods for porous GaN distributed Bragg reflectors," J. Appl. Phys. **139**(13), (2026).

[31] R. Krishna, "Diffusion in porous crystalline materials," Chem. Soc. Rev. **41**(8), 3099–3118 (2012).

[32] T.R. Harris-Lee, T. Turvey, G. Jayamaha, M. Kang, F. Marken, A.L. Johnson, J. Zhang, and C.L. Bentley, "Optimizing Amorphous Molybdenum Sulfide Thin Film Electrocatalysts: Trade-Off between Specific Activity and Microscopic Porosity," ACS Appl. Mater. Interfaces **16**(26), 33620–33632 (2024).

[33] Y. Grishchenko, J. Dawson, S. Ghosh, A. Gundimeda, B.F. Spiridon, N.L. Raveendran, R.A. Oliver, S. Kar-Narayan, and Y. Calahorra, "Magnetic properties of nickel electrodeposited on porous GaN substrates with infiltrated and laminated connectivity," J. Magn. Magn. Mater. **580**, 170877 (2023).

[34] R.A. Oliver, M.J. Kappers, J. Sumner, R. Datta, and C.J. Humphreys, "Highlighting threading dislocations in MOVPE-grown GaN using an in situ treatment with SiH4 and NH3," J. Cryst. Growth **289**(2), 506–514 (2006).

[35] B. Thornley, T.R. Harris-Lee, M.J. Kappers, S.M. Fairclough, and R.A. Oliver, "Improving feature resolution and pore back effect in focused ion beam tomography of porous GaN thin films," ArXiv Preprint, arXiv:2607.07273 (2026).

[36] I.A. Reimers, I. V Safonov, and I. V Yakimchuk, "Segmentation of 3D FIB-SEM data with pore-back effect," J. Phys. Conf. Ser. **1368**(3), 032015 (2019).